\documentstyle[11pt,newpasp,twoside,psfig]{article}
\index{gamma-rays}
\index{binaries: X-ray}
\index{pulsar wind nebulae}
\index{supernova remnants}
\index{unidentified gamma-ray sources}

\def\edcomment#1{\iffalse\marginpar{\raggedright\sl#1\/}\else\relax\fi}
\reversemarginpar

\def\C{\v{C}erenkov } 

\begin{document}
\title{Ground-Based Gamma-Ray Detection of High Energy Galactic Sources: An Update}
 \author{G.P. Rowell}
\affil{Max Planck Institut f\"ur Kernphysik, D-69029 Heidelberg, Germany}

\begin{abstract}
I review the present status of ground-based $\gamma$-ray astronomy, 
concentrating on the population of galactic TeV sources. A number of 
new telescope systems are now being completed, and promise to yield
exciting new discoveries, expanding rapidly the number of sources.
The TeV galactic sources today include a number
of plerions, shell-type SNR, an X-ray binary, and also one unidentified
candidate. Their present status, and our understanding of their TeV $\gamma$-ray emission
processes are summarised and some motivation driving development of the
field is outlined.
\end{abstract}

\section{Introduction}

Ground-based $\gamma$-ray astronomy (at energies $E>100$ GeV) has been developing 
rapidly in the last 10 years. The advent of a number of {\em next-generation} instruments
in 2003/2004 promises exciting
new results and answers to some long-standing questions. Strongly motivating the
development of new detectors and telescopes, are the questions surrounding the 
acceleration of particles to multi-TeV energies in many galactic environments.
It has been known for some time that various evolutionary paths following a 
supernova explosion can lead to the production of high energy cosmic ray (CR) electrons
and hadrons to TeV energies and above, and
with them, $\gamma$-radiation as the most accessible {\em tracer} of this process. 
The main motivation driving the search for galactic TeV sources has been
concerned with the origin of these CR. In particular the shell type SNR have long been
thought responsible for the hadronic CR up to the 'knee' energy $E\sim 10^{15}$eV.
From the first convincing discovery of TeV $\gamma$-rays from the Crab plerion
$\sim$20 years ago (Weekes et al. 1989), galactic objects now comprise a large fraction of 
catalogued TeV $\gamma$-ray emitters.

In this review, I will outline the present and future developments in 
ground-based $\gamma$-ray astronomy, emphasising the imaging atmospheric \C technique, 
and also review the various galactic objects
identified as sources of TeV $\gamma$-rays. Some present and future scientific questions are also
posed. Other reviews of the field, some including summaries of experiments using other techniques
to detect TeV $\gamma$-rays may be found in Aharonian 1999, Weekes 2000, Rowell 2001 and Ong 2003. 

\section{Ground-Based $\gamma$-ray Astronomy: The Imaging Technique}

Ground-based $\gamma$-ray detection relies on the sampling of extensive air showers (EAS) through 
their copious \C photon yield.  EAS comprise the secondary cascade particles (mostly $e^\pm$) 
generated as a primary $\gamma$-ray
interacts with the Earth's atmosphere, and the \C photons carry specific information about EAS development. 
A major feature is the huge effective
collection area $>10^9$ cm$^2$ afforded by the wide area over which the \C photons are distributed over the
ground.
The most powerful technique involves viewing the angular resolution of these \C photons. This
can yield accurate information concerning
the primary $\gamma$-ray energy and direction of origin. Such
telescopes employ large segmented mirrors with focal plane arrays or cameras of fast (ns resolution) photomultiplier tubes. 
These cameras permit parametrisation of \C images, usually as an elliptical profile such that
the major axis aligns with the direction of origin of the primary $\gamma$-ray. Single views of each
image provide an event-by-event angular resolution of $\geq$0.15$^\circ$, for typical pixel granularities
of order $0.15^\circ$ to $0.25^\circ$. 
Imaging cameras have also achieved quite large fields of view (FoV) 
of order 5$^\circ$, permitting survey observations.
The detection of TeV $\gamma$-rays must be made however, against the dominating background of CR.
CR also produce EAS, and hence \C images that are recorded by telescopes. Due to intrinsic differences in the
development of EAS for $\gamma$-rays and CR (hadronic interactions are present in the latter), \C
images for CR tend be more irregular in shape. Furthermore CRs also arrive isotropically, reflecting their
randomised trajectories through interstellar and intergalactic magnetic fields.   
The imaging technique thus provides a powerful means to preferentially select '$\gamma$-ray-like' events via the use
of directional and shape cuts. 
Following the pioneering performance of the Whipple\footnote{\scriptsize http://veritas.sao.arizona.edu/VERITAS\_whipple.html} 
telescope, other major systems such as CANGAROO\footnote{\scriptsize http://icrhp9.icrr.u-tokyo.ac.jp/index.html}, 
CAT\footnote{\scriptsize http://lpnp90.in2p3.fr/\~{}cat/index.html} and 
HEGRA\footnote{\scriptsize http://www-hegra.desy.de/hegra/} 
attained similar, or slightly improved performance, albeit at differing energy thresholds, 
$E\sim 0.25$ to $\sim$ few TeV, owing to various mirror areas and other factors. 

A further, significant improvement in performance is attained from multiple or stereo views of the same \C image. 
By taking advantage of the uncorrelated nature of EAS image fluctuations, systems
of telescopes, separated by $\sim 100$m can achieve an angular resolution proportional to 
$\sim 1/\sqrt{n}$ for $n$ views of 
the same image (Hofmann et al. 1999). For the same reasons, CR background rejection based 
on \C image shape also improves significantly. Typical background rejection fractions can exceed 99.99\% 
after all cuts applicable to a pointlike source. 
The recently decommissioned HEGRA IACT-System (P\"uhlhofer et al. 2003), 
consisting of 5x4 metre diameter telescopes demonstrated clearly the power of stereoscopy coupled with the use of large
FoV cameras. Other benefits of the stereoscopic technique include much improved energy resolution and also a 
reduction in systematics across the FoV, easing the task of performing surveys for new sources.

\section{Galactic Sources: A Status Report}

To date a total of 8 galactic candidates, listed in table~\ref{tab:sourcelist}, are now observed as TeV
$\gamma$-ray sources. Of those listed, only the Crab has been confirmed unambiguously by completely independent groups.
I confine my list here to those sources appearing in the recent refereed literature
($>$1995) and seen at a statistical significance $>4.5\sigma$.
For the most recent results the reader is directed to proceedings of the 28th International Cosmic Ray Conference in 
Japan\footnote{\scriptsize 28th ICRC Japan http://www.icrr.u-tokyo.ac.jp/icrc2003/}, which suggests TeV emission from 
a few more candidates.
%
%
\begin{table}[h]
 \centering
 \begin{tabular}{cccccc} \hline
  Name            & dist  & flux$^a$ & flux$^b$ & signif.$^c$  & Confirm \\
                  & [kpc] &          &          & [$\sigma$]   &         \\ \hline
  \multicolumn{6}{c}{------ Plerions ------}\\
  Crab            &   1.7 &  1.00    &  17.5    & $>50$        & Y \\  
  PSR B1706$-$44  &   1.8 &  1.14    &  20.0    & 12.0         & Y \\
  Vela            &   0.5 &  0.41    &   7.3    & 5.8          & N \\ 
  \multicolumn{6}{c}{------ Shell Type SNR ------}\\
  SN1006          &   1.8 &  0.49    &   8.5    & 7.7          & N\\   
  RXJ1713.7$-$394 &   6.0 &  0.54    &   9.5    & 14.3         & N \\   
  Cas-A           &   3.4 &  0.03    &   0.55   & 4.9          & N \\  
  \multicolumn{6}{c}{------ XRB/Microquasars ------}\\
  Cen X-3         &  $>$5 &  0.64    &  11.2    & 4.5          & N \\ 
  \multicolumn{6}{c}{------ Unidentified ------}\\
  TeV J2032+4130  &     ? &  0.03    &   0.59   & 7.1          & N \\  \hline
  \multicolumn{6}{l}{\scriptsize a: Flux in Crab units  ($E>1$ TeV)}\\
  \multicolumn{6}{l}{\scriptsize b: Flux ($E>1$ TeV) $\times 10^{-12}$ ph cm$^{-2}$ s$^{-1}$}\\
  \multicolumn{6}{l}{\scriptsize c: Statistical significance reported so far (refereed \& unrefereed literature).}\\
 \end{tabular}
 \caption{Summary of galactic TeV sources. Confirmation is defined as postive detection by fully independent
          groups.} 
 \label{tab:sourcelist}
\end{table}
Below I give a brief summary of each listed source, centering on our understanding of their TeV emission 
processes.

\noindent {\em \bf Crab:} The Crab was the first source confirmed as a TeV emitter by the Whipple collaboration, 
and is now clearly detected by all groups active in the field. It is considered the 'standard candle' in the field.
The Crab flux has proved to be steady within instrument systematics, and exhibits an energy spectrum fitted by a pure power 
law over a remarkably large energy range 0.2 to 70 TeV (see for e.g. Horns et al 2003).
The Crab is the most intensely studied pulsar/plerion and the unpulsed TeV flux is universally considered to arise from the 
inverse-Compton (IC) scattering of local (soft \& synchrotron) photons
by the shocked wind electrons filling the Crab plerion. These same electrons give rise to the very 
intense broad-band synchrotron spectrum extending from radio to X-ray energies. At the high energy end however
the contribution of accelerated hadrons which lead to $\gamma$ emission via $\pi^\circ$ decay
could also become significant (Aharonian \& Atoyan 1996a). This rather hard 'hadronic
component' may be required to explain the unbroken continuation of the Crab spectrum to $E>50$ TeV. An IC cutoff would normally
be expected due to electron cooling and the Klein-Nishina reduction in IC cross section.\\
{\em \bf PSR B1706$-$44:} Steady TeV emission from this source was reported by CANGAROO and confirmed by the
Durham group (Kifune 1995; Chadwick et al. 1998). As for the Crab, the plerionic shocked pulsar wind could 
give rise to the unpulsed TeV emission, which is reportedly at similar levels to the Crab TeV flux. However, 
the very weak synchrotron
X-ray flux, only recently seen by {\it Chandra}  (Gotthelf et al. 2002) leads to a strong underprediction of the IC
TeV for 'reasonable' values of the magnetic field $B\sim 15\mu$G. One possible way out is for the high energy electrons
to escape quickly to regions of lower $B$ (Aharonian et al. 1997).\\
{\em \bf Vela:} This is another plerion, whose unpulsed TeV emission has been interpreted in the electronic IC framework. 
Complicating matters somewhat, the CANGAROO discovery
found the TeV emission 0.13$^\circ$ offset from the pulsar position (Yoshikoshi et al. 1997), not far from its 
supposed birthplace. It is possible that a 'relic' electron wind
may be responsible for the plerionic synchrotron X-ray and IC TeV emission although this implies very low $B$ fields
(Harding et al. 1997). An alternative hadronic picture is put forward by Aharonian 1999 as well as the unshocked electronic 
wind scenario of Aharonian \& Bogovalov (2003).\\
{\em \bf SN1006:} This historical young shell type SNR expanding into a relatively sparse environment exhibits a predominantly
featureless X-ray spectrum, considered to be synchrotron emission. The bi-polar morphology of SN1006 in X-rays may
trace out regions of more efficient electron injection. TeV emission has been seen from the
slightly brighter (in X-rays) NE rim (Tanimori et al. 1998), and was initally attributed to the electronic IC process.
However, as for PSR B1706$-$44 and Vela, very low $B<10 \mu$G fields are required 
to match the CANGAROO flux. It also appears that the hadronic component may play a significant role
(Aharonian \& Atoyan 1999; Berezkho et al. 2002), particularly given the recent high-resolution 
results of {\em Chandra} which favour higher $B>10 \mu$G fields in the post-shocked regions (Bamba et al. 2003).\\  
{\em \bf RXJ1713.7$-$394:} This older shell-type SNR appears to show TeV emission at its NW rim (Enomoto et al. 2002), 
not far from a neighbouring molecular cloud with high ambient matter density. There is also
strong non-thermal X-ray emission from this area. Interpretation 
was initially centred on the electronic IC component but more recently re-evaluation was made in strong favour of the
hadronic origin for the accelerated particles. As for SN1006, the high resolution filamentary structure seen by
{\em Chandra}, pointing to a high $B$ field would certainly favour the hadronic component as discussed by 
Uchiyama et al. 2003.\\
{\em \bf Cas A:} The youngest shell-type SNR, Cas-A is one of the strongest radio sources. The HEGRA detection
of TeV emission was made using a very deep exposure of $>200$ hrs (Aharonian et al. 2001a). The TeV flux
appears to favour a hadronic interpretation since the very high magnetic field in Cas-A, few$\sim$mG, quenches strongly any electronic
component (IC and Bremsstrahlung) at $E>1$ TeV (Atoyan et al. 2000; Berezhko et al. 2003). More detailed spectral information
at $E>5$TeV should solve this issue.\\
{\em \bf Cen X-3:} This high mass X-ray binary contains a 4.8 s pulsar in a 2.1 day orbit around an O-type supergiant
and has also been seen by EGRET at low GeV energies. Neither the GeV or TeV
fluxes appear modulated with the pulsar orbital period (Chadwick et al. 2000). The TeV production could come from a 
beam of relativistic hadrons interacting with ejected matter clumps from the companion star (Aharonian \& Atoyan 1996b).
Recent reanalysis of the TeV data does suggest some modulation of the TeV emission with the pulsar half-period 
from one night (Atoyan et al. 2002) for higher energy events. In general, the lack of modulation could 
point to a large scale source region, quite distant from the neutron star and Atoyan et al. also consider 
electronic IC components and possible ways to discriminate them from the hadronic scenario.\\
{\em \bf TeV~J2032+4130:} Discovered seredipitously in HEGRA IACT-System data (Aharonian et al. 2002; Rowell et al. 2003), 
TeV~J2032+4130 is at present the only unidentified
TeV source. Despite the fact that no multiwavelength counterpart is identified, its location within the
extremely dense OB association Cygnus~OB2 cluster could point to particle acceleration in such environments. Alternative
explanations invoke a jet-powered microquasar scenario, for example a link to nearby Cygnus X-3, the EGRET source
3E2033+4118, or even an unseen plerion. 
The lack of an X-ray counterpart could suggest a hadronic source and deeper X-ray observations of the Cygnus OB2 region are 
in the pipeline (Butt et al. 2003). The discovery of TeV~J2032+4130 is in fact a perfect example of the survey capabilities
of present ground-based instruments, particularly those utilising large fields of view and achieving arc-minute source
location error boxes. 
 
Overall, the interpretion of the galactic TeV sources has invoked both the hadronic and electronic models. 
Except for the Crab, which is quite well understood in terms of the electronic synchrotron/IC framework,
solid discrimination between the hadronic and electronic framework requires more accurate TeV spectral information
covering a broader energy range, coupled with multiwavelength studies. In some cases the high resolution results 
from {\em Chandra} are certainly helping to constrain the electronic component.
So far what is lacking is clear proof that the TeV galactic objects are capable of accelerating hadrons to multi-TeV
energies. Further improvements in sensitivity and
energy coverage for ground-based instruments have therefore been sought after during the last 5 years or so. 
In particular as well as aiming for a decrease in energy threshold to $E<300$ GeV, it is now clear that sensitivity 
improvements at the high energy end $E>10$ TeV are are also deemed vital to separating the hadronic from electronic 
emission processes. 

\section{New Ground-Based Instruments}
\begin{figure}[t]
 \psfig{file=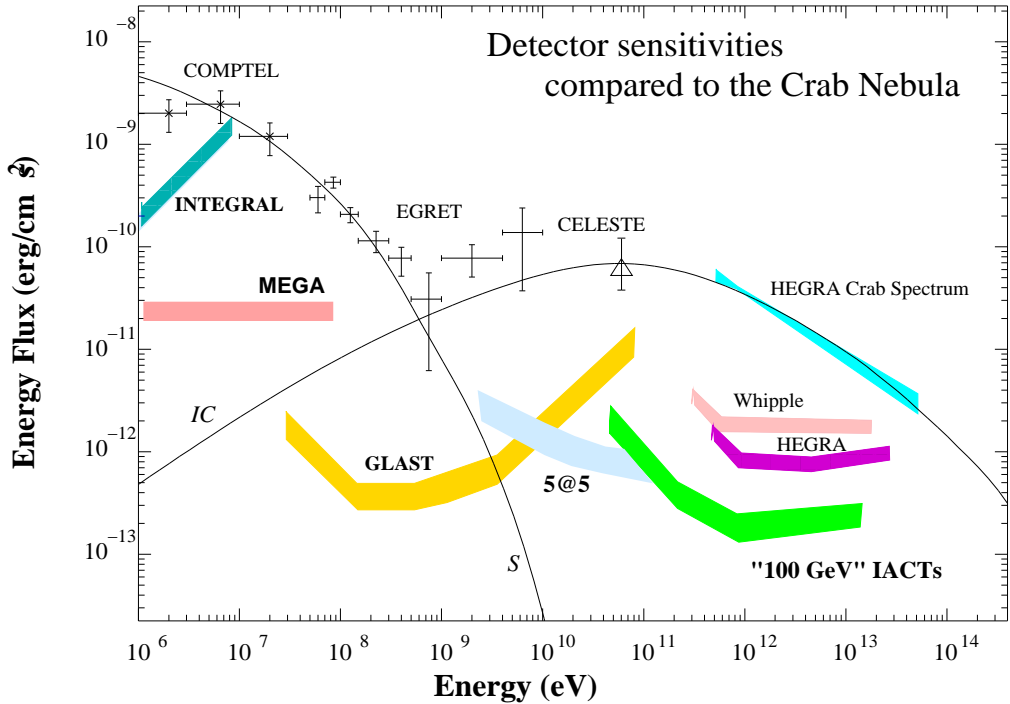,width=\textwidth}
 \caption{Comparison of detector sensitivies against the Crab broad-band spectrum. '100 GeV IACTs' refer to 'Next Generation''
  instruments described in the text (H.E.S.S., CANGAROO III, VERITAS and MAGIC). For the ground-based instruments a 
  50 hour (5 $\sigma$) detection limit is required. 
  For GLAST, INTEGRAL, 
  and MEGA a 10$^6$~sec integration time is specified.}
 \label{fig:sensitivities}
\end{figure}

The best ground-based systems operating up until 2002 achieved an angular resolution, energy resolution, energy threshold, 
and sensitivity of $\la 0.1^\circ$, $\leq$15\%, $\sim$250 GeV and 
$\sim1\times10^{-12}$ erg cm$^{-2}$ s$^{-1}$ ($E>$1 TeV, 50 hours, 
5$\sigma$) respectively. This flux sensitivity amounts to $\sim$5 to 10$\sigma$ / $\sqrt{hr}$ on the Crab.
The so-called {\em next generation} instruments in ground-based $\gamma$-ray astronomy aim to realise
roughly one order of magnitude improvement in sensitivity and reduction in energy threshold over present instruments.
Four primary projects are now coming online. The 
H.E.S.S.\footnote{\scriptsize http://www.mpi-hd.mpg.de/hfm/HESS/HESS.html}, 
VERITAS\footnote{\scriptsize http://veritas.sao.arizona.edu/veritas/index.html} 
and CANGAROO III\footnote{\scriptsize http://icrhp9.icrr.u-tokyo.ac.jp/c-ii.html} projects
are employing the stereoscopic imaging technique. Data analysis techniques will be in general similar to
that used by present instruments, in particular the stereoscopic aspects developed with the HEGRA IACT-System.
These systems consist of arrays of $\ge 4$ telescopes each with $\sim$100 m$^2$ segmented mirrors and imaging cameras of 
$\ge$500 pixels subtending $\ge 3^\circ$ fields of view. All are expected to achieve energy thresholds in the range
50 to 200 GeV. The MAGIC\footnote{\scriptsize http://hegra1.mppmu.mpg.de/MAGICWeb/}
project comprises a single, very large telescope of mirror area ($>$200 m$^2$) in an effort to achieve an energy
threshold $E<50$ GeV. Extra telescopes may be added to form a stereoscopic system. Figure~\ref{fig:sensitivities}
compares the sensitivities of present and next-generation (and beyond) ground-based instruments along with the
next space-based detectors in the high energy regime GLAST, INTEGRAL and MEGA.

As of October 2003, H.E.S.S. and CANGAROO-III have a number of telescopes running. They are expecting 
completion of their full 4-telescope arrays by 2004. VERITAS has just begun partial operation
of their prototype telescope and MAGIC is also nearing completion, already taking engineering data.
Both H.E.S.S. and CANGAROO are situated in the southern hemisphere and as such are perfectly postitioned to survey 
the Galactic plane for new sources. Searching for high energy counterparts to the numerous EGRET unidentified 
sources (Hartman et al. 1999) is a prime motivation.
The angular resolution and FoV of H.E.S.S. for example is expected
to survey the inner plane ($l\pm 30^\circ$, $b\pm 2.5^\circ$) at the 0.1 Crab level in less than 200 hours observation time. 
H.E.S.S. is also presently devoting large effort to confirmation of the all of the southern TeV sources listed
above, all of which require confirmation at solid significance levels similar to early results on the Crab for 
full acceptance in the field.

Looking further into the future, already the expansion of H.E.S.S. and the like is under full consideration. Several options 
ere now being considered as to how to improve the energy coverage and sensitivities even further. One such option
given serious thought is the placement of a \C telescope at high mountain altitude $>4000$ m to approach the lowest
possible threshold allowed by the technique $E\sim 5$ GeV. The '5@5' proposal (Aharonian et al. 2001b) considers
placing a H.E.S.S.-like system (slightly bigger telescopes) at the ALMA site in Chile (5000 m). Such a system could
yield significant detections of Vela in just seconds, allowing for the first time $\gamma$-ray pulse timing activities. 
The follow-up of $\gamma$-ray bursts is also a major consideration.

\acknowledgments
I thank the organisers of IAU Symposium 218 for the invitation.

\end{document}